\documentclass[twocolumn,pre,superscriptaddress,amsmath,amssymb,longbibliography]{revtex4-1}

\usepackage{graphicx} % Required for inserting images
\usepackage{dcolumn}
\usepackage{bm}

\usepackage{float}
\usepackage{color}
\begin{document}

\title{Multihyperuniform Particle Composites Inspired by Avian Photoreceptor Patterns for Optical Applications}
\author{David Keeney}
\affiliation {Materials Science and Engineering, Arizona State University, Tempe, AZ 85287}
\author{Wenlong Shi}
\affiliation {Materials Science and Engineering, Arizona State University, Tempe, AZ 85287}
\author{Rohit Thomas}
\affiliation {Arizona College Preparatory High School, Chandler, AZ 85249}
\author{Yang Jiao}
\affiliation {Materials Science and Engineering, Arizona State University, Tempe, AZ 85287} \affiliation{Department of Physics, Arizona State University, Tempe, AZ 85287}
\date{\today}

\begin{abstract}
Hyperuniform materials, characterized by anomalously suppressed long-wavelength density fluctuations, exhibit unique optical and photonic properties distinct from both crystalline and random media. While most prior studies have focused on single-species systems, we investigate the broader class of \textit{multihyperuniform} systems inspired by biological photoreceptor mosaics. Using particle-based models with varying species number, size ratios, and interaction competition, we demonstrate that multispecies mixtures can achieve robust and stealthy hyperuniform configurations, even in highly disordered states. We further show how these configurations can be mapped to multifunctional composites with tailored optical responses, including isotropic structural coloration, enhanced absorption, and engineered dielectric properties that facilitate transmission while suppressing scattering. Our results highlight multihyperuniformity as a generalizable design principle for multifunctional disordered photonic materials, opening avenues for robust, tunable, and scalable optical applications.
\end{abstract}

\maketitle

\section{Introduction}

%\textcolor{blue}{Introduce hyperuniform materials, the concept, brief definition, and novel properties especially related to optics and photonics applications.} 

%\textcolor{blue}{Mention the preponderance of previous work focuses on binary systems, and very few studies on multihyperuniform systems, cite the existing papers.} A \textit{multihyperuniform} many-body system \cite{Ji14, lomba2020minimal, chen2021multihyperuniform, hong2024nanowire, frusawa2025theoretical}. Talk about existing optical applications of these systems.

%\textcolor{blue}{Mention this work focuses on generalization multi-hyperuniform systems inspired by avian receptors. Explore the competition of short-range vs. long-range interactions, across multiple number of species, number ratios, etc. It is easier to generalize to 3D dimensions, on or off lattices.}

%\textcolor{blue}{Mention converting the configurations to composites with optical applications: for observing of certain light yet facilitating prorogation of others.}

%\textcolor{blue}{breifly outline the rest of the paper.}

The concept of hyperuniformity has emerged as a unifying framework to describe a remarkable class of disordered many-body systems and materials in which large-scale density fluctuations are anomalously suppressed \cite{To03, To18a}. A \textit{hyperuniform} many-particle system, disordered or not, is one in which the number variance of particle centers within a spherical observation window of radius $R$ grows slower than the observation window volume $R^d$ in $d$-dimensional space for large $R$ \cite{To03,To18a}, i.e.,
\begin{equation}
\lim_{R\rightarrow\infty}\sigma_N^2(R)/v_1(R) = 0
\end{equation}
where $\sigma_N^2(R)\equiv\langle N(R)^2\rangle - \langle N(R)\rangle^2$ is the variance in the number of particle centers (brackets indicate the ensemble average) and $v_1(R) \sim R^d$ is the volume of a $d$-dimensional sphere or radius $R$. Equivalently, a system is hyperuniform if its structure factor $S(\mathbf{k})$ (defined in Sec. II) tends to zero as the wavenumber $k\equiv|\mathbf{k}|$ tends to zero, i.e.,
\begin{equation}
    \lim_{|\mathbf{k}|\rightarrow0}S(\mathbf{k})=0.
\end{equation}
The small-$|{\bf k}|$ scaling behavior of $S({\bf k}) \sim |{\bf k}|^\alpha$ determines the large-$R$
asymptotic behavior of $\sigma_N^2(R)$, based on which all hyperuniform
systems, disordered or not, can be categorized into three classes:
$\sigma_N^2(R) \sim R^{d-1}$ for $\alpha>1$ (class I); $\sigma_N^2(R)
\sim R^{d-1}\ln(R)$ for $\alpha=1$ (class II); and $\sigma_N^2(R)
\sim R^{d-\alpha}$ for $0<\alpha<1$ (class III) \cite{To18a}. Consequently, hyperuniform systems encompass all crystals and quasicrystals \cite{To03, To18a}. Hyperuniformity has also been identified in a wide spectrum of physical \cite{ref4, ref5, ref6, ref7, ref16, ref17, ref18, ref19, ref20,
ref21, ref22, ref23, salvalaglio2020hyperuniform, hexner2017noise, hexner2017enhanced, weijs2017mixing,
lei2019nonequilibrium, lei2019random, ref8, ref9, ref10, ref11,
ref12, ref13, ref14, ref15, ref24, ref25, sanchez2023disordered}, material \cite{ref28, ref29, ref30, Ge19, sakai2022quantum, Zh20, Ch21, PhysRevB.103.224102, Zh21, nanotube, zhang2023approach, chen2021multihyperuniform, chen2025anomalous, shi2025three, wang2025hyperuniform, zhong2025modeling} and biological \cite{ref26, ref27, ge2023hidden, liu2024universal, tang2024tunablehyper} systems. 

The unique combination of local disorder and complete suppression of global density fluctuations endows these systems with unexpected novel physical properties, including wave propagation characteristics \cite{ref31, ref32, ref33, scattering, granchi2022near, park2021hearing, klatt2022wave, tavakoli2022over, cheron2022wave, yu2021engineered, li2018biological}, thermal, electrical and diffusive transport properties \cite{ref34, torquato2021diffusion, maher2022characterization}, mechanical properties \cite{ref35, puig2022anisotropic} as well as optimal multifunctional characteristics \cite{ref36, kim2020multifunctional, torquato2022extraordinary}. In optics and photonics, hyperuniform point patterns and composites have been shown to support isotropic photonic band gaps, angle-independent structural coloration, suppressed scattering, and enhanced transmission. These properties position hyperuniform materials as promising platforms for broadband photonic devices, structural color coatings, transparent conductors, and low-loss metamaterials.

While most prior work has focused on single-species hyperuniform systems, in which one type of particles are arranged to achieve hyperuniform order, comparatively few studies have explored the richer landscape of \textit{multihyperuniform} many-body systems \cite{Ji14, lomba2020minimal, chen2021multihyperuniform, hong2024nanowire, frusawa2025theoretical}, in which each individual species in a multi-species system exhibits hyperuniformity. These investigations demonstrate that multiple species can collectively enforce hyperuniformity while allowing greater tunability in local correlations, spectral characteristics, and disorder tolerance. Existing applications of such multihyperuniform systems include optical nano-wire arrays as artificial photoreceptors inspired by avian photoreceptors \cite{hong2024nanowire}, 

In this work, we focus on the study of multihyperuniform systems inspired by the organization of avian photoreceptor cells, which provide a natural example of a five-species multihyperuniform mosaic \cite{Ji14}. We explore the competition between short-range and long-range interactions in generating hyperuniformity across multiple species, number ratios, and density conditions. This framework provides a systematic route for understanding how increasing the number of species affects local packing, long-range correlations, and the emergence of stealthy multihyperuniform states. Moreover, we demonstrate how these multihyperuniform configurations can be mapped onto composite materials with tailored optical functionalities. Such composites can be engineered to selectively absorb or scatter light at targeted wavelengths while simultaneously facilitating efficient propagation at others. The multifunctional design enabled by multihyperuniformity opens the door to composites that integrate isotropic coloration, controlled absorption, and engineered dielectric responses, offering a versatile platform for next-generation photonic applications.

The rest of the paper is organized as follows: In Sec. II, we provide necessary definitions of hyperuniformity and related concepts. In Sec. III, we describe the cherry-pit packing model for generating multihyperuniform many-particle configurations. In Sec. IV, we present our computational investigations of binary and multi-species systems, focusing on the effects of competition between two-scale interactions, particle number ratio, and number of species. We also demonstrate multi-functional applications of the resulting multihyperuniform configurations. In Sec. V, we provide concluding remarks.

%A special subset of class-I hyperuniform systems possess a zero structure factor for a range of wavenumbers $k = |{\bf k}|$ around the origin, e.g., $S(k) = 0$ for $k<K^*$ (excluding the forward scattering), which are referred to as stealthy hyperuniform systems. Stealthy hyperuniform systems include all crystals and certain special disordered systems, which are characterized by a parameter $\chi \in (0, 0.5)$ (defined in Sec. II) reflecting the fraction of the constrained degrees of freedom in the system \cite{To15, Zh15a, Zh15b, Zh17}. Increasing $\chi$ also leads to an increase of the local order in the system. 

%the hyperuniform correlations are distinctly different from those in conventional correlated disordered systems, such as equilibrium or nonequilibrium hard-sphere systems \cite{torquato2010jammed}. In the latter, an increase of local order, resulting from mutual exclusion effects, can be achieved via increasing the packing fraction of the system. However, the hard-particle systems lack the quasi-long-range correlations that exist in hyperuniform systems and give rise to complete suppression of density fluctuations in the infinite wavelength limit. The sole exception is the special randomly jammed packing state, which was shown to be hyperuniform \cite{donev2005unexpected, maher2023hyperuniformity}.

\section{Definitions}

A many-particle configuration in $d$-dimensional Euclidean space $\mathbb{R}^d$ is completely characterized by an infinite set of $n$-point correlation functions $\rho_n(\mathbf{r}_1,\dots,\mathbf{r}_n)$, which are proportional to the probability of finding $n$ points (representing the centers of particles) at the positions $\mathbf{r}_1,\dots,\mathbf{r}_n$ \cite{torquato2002random}.
For statistically homogeneous systems, $\rho_1(\mathbf{r}_1)=\rho$, and $\rho_2(\mathbf{r}_1,\mathbf{r}_2)=\rho^2 g_2(\mathbf{r})$, where $\mathbf{r}=\mathbf{r}_1-\mathbf{r}_2$, and $g_2(\mathbf{r})$ is the pair correlation function.
For statistically isotropic systems, $g_2(\mathbf{r})=g_2(r)$, where $r=|{\bf r}|$.
The associated structure factor $S(\mathbf{k})$ is defined as
\begin{equation}
    S(\mathbf{k})=1+\rho\Tilde{h}(\mathbf{k})
\end{equation}
where $\Tilde{h}(\mathbf{k})$ is the Fourier transform of the total correlation function $h(\mathbf{r})=g_2(\mathbf{r})-1$, and ${\bf k}$ is the wave vector.

For a single periodic point configuration with $N$ particles at positions $\mathbf{r}^N = (\mathbf{r}_1,\dots,\mathbf{r}_N)$ within a fundamental cell $F$ of a lattice $\Lambda$, the scattering intensity $\mathbb{S}(\mathbf{k})$ is given by
\begin{equation}\label{eq:Skcomp}
    \mathbb{S}(\mathbf{k}) = \frac{|\sum_{j=1}^N \textrm{exp}(-i\mathbf{k}\cdot\mathbf{r}_j)|^2}{N}.
\end{equation}
In the thermodynamic limit, the scattering intensity of an ensemble of an $N$-particle configurations in $F$ is related to $S(\mathbf{k})$ by
\begin{equation}
    \lim_{N,V_F\rightarrow\infty}\langle \mathbb{S}(\mathbf{k})\rangle = (2\pi)^d \rho \delta(\mathbf{k}) + S(\mathbf{k}),
\end{equation}
where $\rho = N/V_F$, $V_F$ is the volume of the fundamental cell, and $\delta$ is the Dirac delta function \cite{To03}.
For finite-$N$ simulations under periodic boundary conditions, Eq. (\ref{eq:Skcomp}) is used to compute $S(\mathbf{k})$ for all nonzero wave vectors. 

%\subsection{System Types}
%\subsubsection{HIP Systems}
%\subsubsection{Poisson Systems}
%\subsubsection{RSA Systems}
%\subsubsection{Disordered Non-Stealthy Hyperuniform Systems}
%\subsubsection{Disordered Stealthy Hyperuniform Systems}
%\subsection{Voronoi Tessellation}

We mainly consider systems characterized by a structure factor with a radial power law in the vicinity of the origin,
\begin{equation}
S(\mathbf{k})\sim|\mathbf{k}|^{\alpha}\;\textrm{for}\;|\mathbf{k}|\rightarrow0,
\end{equation}
where the exponent $\alpha$ is referred to as the hyperuniformity exponent. For {\it hyperuniform} systems, $\alpha > 0$. A (standard) {\it nonhyperuniform} system has $\alpha = 0$, i.e., S({\bf k}) approaches a non-zero constant in the zero-wavenumber limit. An {\it antihyperuniform} system is one possessing a diverging S({\bf k}) in the zero-wavenumber limit, i.e., with $\alpha < 0$.

For hyperuniform systems, the specific value of $\alpha$ determines large-$R$ scaling behaviors of the number variance \cite{To18a}, according to which all hyperuniform systems can be categorized into three different classes:
\begin{equation}\label{eq:classes}
    \sigma^2_N(R)\sim
    \begin{cases}
    R^{d-1}&\alpha > 1, \textrm{class I}\\
    R^{d-1}\textrm{ln}(R)&\alpha = 1, \textrm{class II}\\
    R^{d-\alpha}&\alpha < 1, \textrm{class III}.
    \end{cases}
\end{equation}
Classes I and III are the strongest and weakest forms of hyperuniformity, respectively. Class I systems include all crystals \cite{To03}, many quasicrystals \cite{Og17}, and certain exotic disordered systems \cite{Za09, Ch18a}. Examples of Class II systems include some quasicrystals \cite{Og17}, perfect glasses \cite{zhang2017classical}, and maximally random jammed packings \cite{Do05, Za11a, Ji11, Za11c, Za11d}. Examples of Class III systems include classical disordered ground states \cite{Za11b}, random organization models \cite{He15}, perfect glasses \cite{zhang2017classical}, and perturbed lattices \cite{Ki18}; see Ref. \cite{To18a} for a more comprehensive list of systems that fall into the three hyperuniformity classes.

%Such classes apply analogously to spectral densities possessing a radial power-law form in the vicinity of the origin, i.e. \cite{Torquato_DisorderHUHet},
%\begin{equation}
%    \spD{\mathbf{k}}\sim|\mathbf{k}|^{\alpha}\;\textrm{for}\;|\mathbf{k}|\rightarrow0.
%\end{equation}
%These classes instead correspond to a large-$R$ scaling behavior \textit{volume-fraction} variance, which considers the volume fraction of a desired phase in an observation window of radius $R$.
%\begin{equation}
%        \sigma^2_V(R)\sim
%    \begin{cases}
%    R^{-(d-1)}&\alpha > 1, \textrm{class I}\\
%    R^{-(d-1)}\textrm{ln}(R)&\alpha = 1, \textrm{class II}\\
%    R^{-(d-\alpha)}&\alpha < 1, \textrm{class III}.
%    \end{cases}
%\end{equation}

Stealthy hyperuniform systems are a special subset of class-I hyperuniform systems possessing a zero structure factor for a range of wavevectors around the origin, i.e.,
\begin{equation}
S({\bf k}) = 0, \quad\text{for}\quad {\bf k} \in \Omega,
\label{eq_stealthy}
\end{equation}
excluding the forward scattering. Stealthy hyperuniform systems include all crystals and certain special disordered systems, which are characterized by a parameter $\chi$ reflecting the fraction of the constrained degrees of freedom in the system \cite{To15, Zh15a, Zh15b, Zh17}, i.e., $\chi = N_\Omega/(N-d)$
where $N_\Omega$ and $N$ are respectively the number of constrained and total degrees of freedom, and $d$ degrees of freedom associated with the trivial overall translation of the entire system are subtracted. In the case of point configurations, it has been shown that increasing $\chi$ leads to an increased degree of local order in the systems \cite{To18a, Ba08}. Disordered stealthy hyperuniform systems have been shown to possess unique optical properties such as isotropic complete photonic band gaps \cite{Fl09, Fl13} and superior electromagnetic wave transmission \cite{torquato2021nonlocal, kim2023effective, kim2023extraordinary}.

A \textit{multihyperuniform} many-body system \cite{Ji14, lomba2020minimal, chen2021multihyperuniform, hong2024nanowire, frusawa2025theoretical} is one that contains $n_s \ge 2$ different components, in which each individual component is hyperuniform, i.e., $S_i(\mathbf{k})\sim|\mathbf{k}|^{\alpha_i}$ for $|\mathbf{k}|\rightarrow0$ ($i = 1, \ldots, n_s$). Equivalently, the number variance $\sigma_{N, i}^2(R)$ for each component $i$ satisfies  $\lim_{R\rightarrow\infty}\sigma_{N, i}^2(R)/v_1(R) = 0$. We note that the individual components may belong to different hyperuniformity classes, thereby opening a rich spectrum of possible structural configurations. Existing examples of multihyperuniform systems include avian photoreceptor patterns \cite{Ji14}, nanowire patterns for artificial photoreceptors \cite{hong2024nanowire} and certain high-entropy alloys \cite{chen2021multihyperuniform}. 
%\textcolor{red}{need to cite the japanese paper on theory}

\section{Cherry-Pit Packing Model}

%\subsection{ to Generate Multihyperuniform Systems}

\begin{figure}[ht]
\begin{center}
$\begin{array}{c}\\
\includegraphics[width=0.45\textwidth]{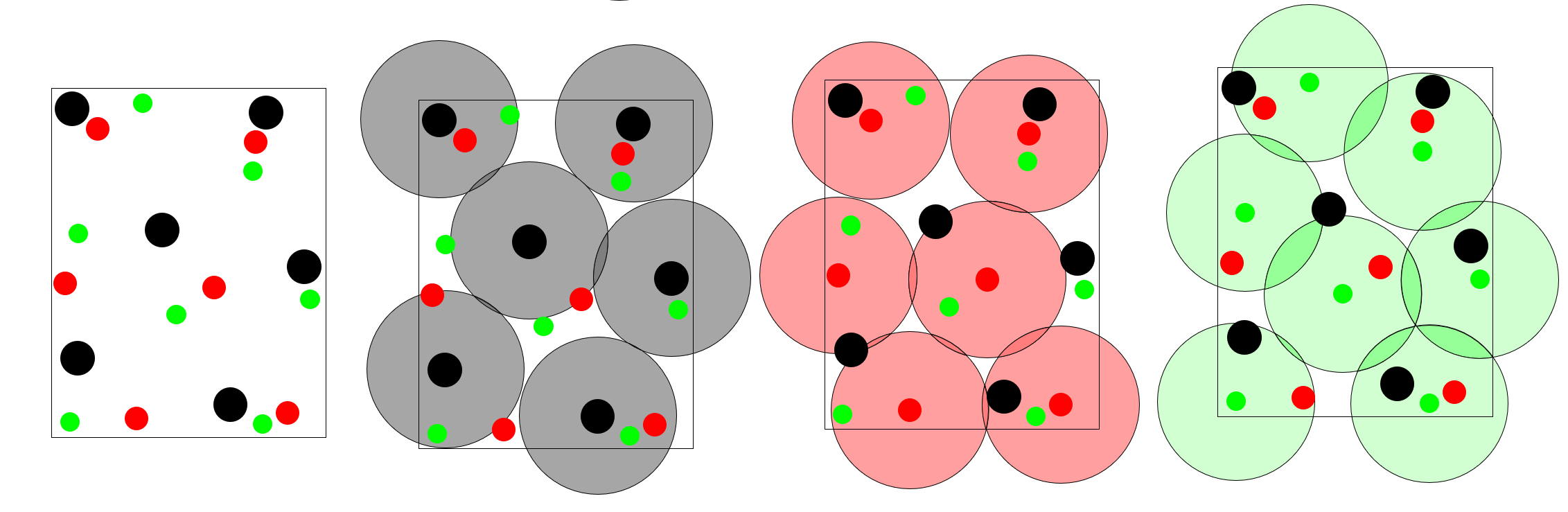}
\end{array}$
\end{center}
\caption{Illustration of the cherry-pit packing model of a 3-species systems, in which short-range hard-core repulsion is between any pair of particles with different radius and long-range soft-shell repulsion is only between particles of the same species.} \label{fig_1}
\end{figure}

We employ a cherry-pit packing model developed in Ref. \cite{Ji14} to generate multihyperuniform many-particle systems, see Fig. \ref{fig_1} for illustration. Following Ref. \cite{Ji14}, we consider the particles possess a long-range soft-shell homotype repulsion (i.e., only existing between particles of the same species) as well as a short-range hard-core exclusion between all particles. The soft-shell repulsion for particles of species $i$ is characterized by an isotropic pair potential 
\begin{equation}
\varepsilon^{(i)}(r) = 
 \begin{cases}
    \frac{\alpha}{\beta+1}(2R^{(i)}_s - r), &\textrm{ for } r \le 2R^{(i)}_s\\
    0, &\textrm{ for } r > 2R^{(i)}_s,
    \end{cases}
\end{equation}
where $\alpha$, $\beta$ and $R^{(i)}_s$ respectively sets the scale, strength and range of the interaction. Here, we choose $\alpha = \beta = 1$ and $R^{(i)}_s = \frac{1}{2}\sqrt{2/(\sqrt{3}\rho^{(i)})}$, where $\rho^{(i)}$ is the number density of particles of species $i$. These parameter values have been shown to robustly lead to hyperuniform configurations for $n_s = 5$ species systems \cite{Ji14}. Moreover, each particle of species $i$ possesses a circular hard core with radius $R_h^{(i)}$, which is gradually increased during the simulation. The packing density $\phi$ of the particles is given by
\begin{equation}
    \phi = \frac{1}{L^2}\sum_{i=1}^{n_s} N_i \pi [R^{(i)}_h]^2 
\end{equation}
where $L$ is the edge length of the square periodic simulation domain. The particles of a specific species $i$ possess the same $R^{(i)}_s$ and $R^{(i)}_h$.   

An initial configuration of a total $N =\sum_i^{n_s} N_i$ particles of $n_s \ge 2$ species with packing density $\phi_0 = 0.05$ is generated using the random sequential addition process, which is followed by Monte Carlo (MC) simulations. At each MC stage, the following steps are carried out:
\begin{itemize}
    \item Hard-core equilibration and growth:  Particles of all species are randomly selected and displaced within a maximal distance   ($~0.25 R_h^{(i)}$) without violating the nonoverlapping constraints. On average, each particle is moved $\mathcal{N} \sim 1000$ time before a growth step, where the hard-core radii $R_h^{(i)}$ of the particles are proportionally increased in size while maintaining the size ratio so that the packing fraction $\phi$ is increased by roughly 1$\%$. Notably during this equilibration and growth stage, long-range soft-shell interactions are not active.
    \item Soft-shell relaxation: If the packing density $\phi$ exceeds a prescribed value $\phi_r$ (i.e., the relaxation density), the long-range homotype soft-shell repulsion is activated. While still subject to the nonoverlapping constraints imposed by the hard cores, the steepest descent method is used to bring the system to its closest local minimum of the potential energy
    \begin{equation}
        \Psi = \sum_{i=1}^{n_s}\sum_{m, n} \varepsilon^{(i)}(r_{mn})
        \label{eq_energy}
    \end{equation}
    where $m$ and $n$ are a pair of particles of species $i$. The energy minimization effectively relaxes and homogenizes the particles of the same species, leading to a strong suppression of density fluctuations in the system.
\end{itemize}

The above steps are repeated until either a prescribed termination density $\phi_f$ or a set maximum number of stages is reached. Statistics of particle configurations including their pair correlation function $g_2(r)$, statistic structure factor $S(k)$, number variance $\sigma_N^2(R)$ and nearest-neighbor distribution $n(r)$ for all species are collected after each MC stage. The definitions of these statistics are provided in Sec. II.

We note that the polydispersity associated with the hard-cores across different species generally suppresses crystallization in the system and can lead to segregation of particles of the same species (with the same $R_h^{(i)}$), especially at higher packing densities. However, homotype soft-shell repulsion effectively spreads out and homogenizes the distribution of the particles of the same species. As pointed out in Ref. \cite{Ji14}, the competition between the hard-core and soft-shell interactions can give rise to the observed multihyperuniformity in the avian photoreceptor patterns. Here, we systematically investigate how this competition depends on the relaxation density $\phi_r$, particle number ratios, and the number of species. We study a wide range of particle numbers $N = 200 \sim 2500$ in the subsequent simulations and confirm that the our findings are not affected by the choice of $N$.

\section {results}

\subsection {Effects of Long-Range Soft-Shell Interactions}

\begin{figure}[ht]
\begin{center}
$\begin{array}{c}\\
\includegraphics[width=0.35\textwidth]{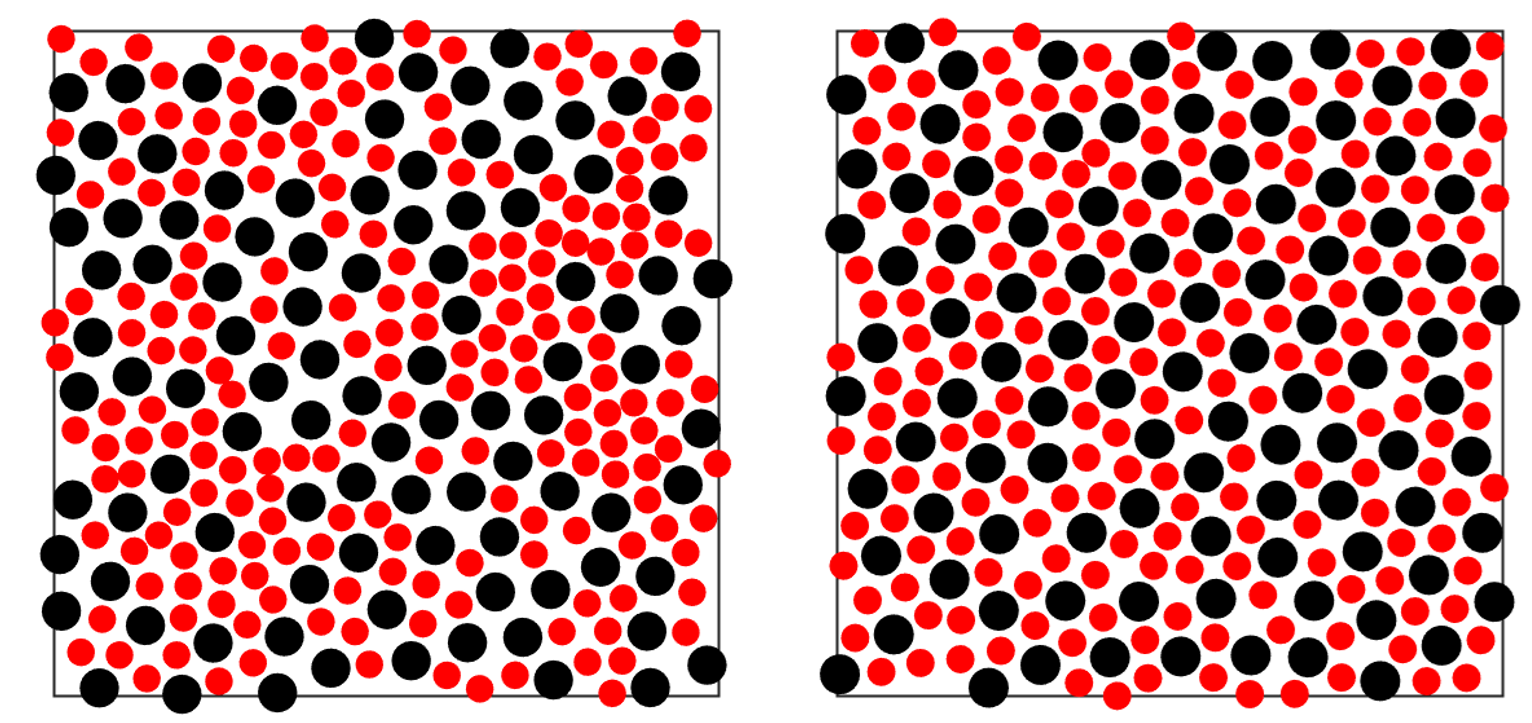}
\end{array}$
\end{center}
\caption{A binary system with $N=300$, a 1:1 number ratio at $\phi_f \approx 0.7$. Size ratio of large (black) to small (red) particles is 2:1. Left panel: Hard-core interactions lead to segregation of the large and small particles into clusters of same type. Right panel: Soft-shell interactions spread out the particles, leading to a more uniform distribution.} \label{Binary_2m_per_stage}
\end{figure}

In our simulations, at each MC stage, the system is first equilibrated with respect to the hard-core interactions. If the packing density $\phi$ exceeds a prescribed relaxation density $\phi_r$, the homotype soft-shell repulsion is activated and the particles are re-organized to minimize the total energy (\ref{eq_energy}). Figure \ref{Binary_2m_per_stage} illustrates the effects of the homotype soft-shell repulsion in a binary system: Before the interactions are activated, the particles of the same type tend to segregate into clusters (left panel). After the system is relaxed to minimize the total energy due to soft shell repulsion, particles of both species are uniformly spread out (right panel).

\begin{figure}[ht]
\begin{center}
$\begin{array}{c}\\
\includegraphics[width=0.475\textwidth]{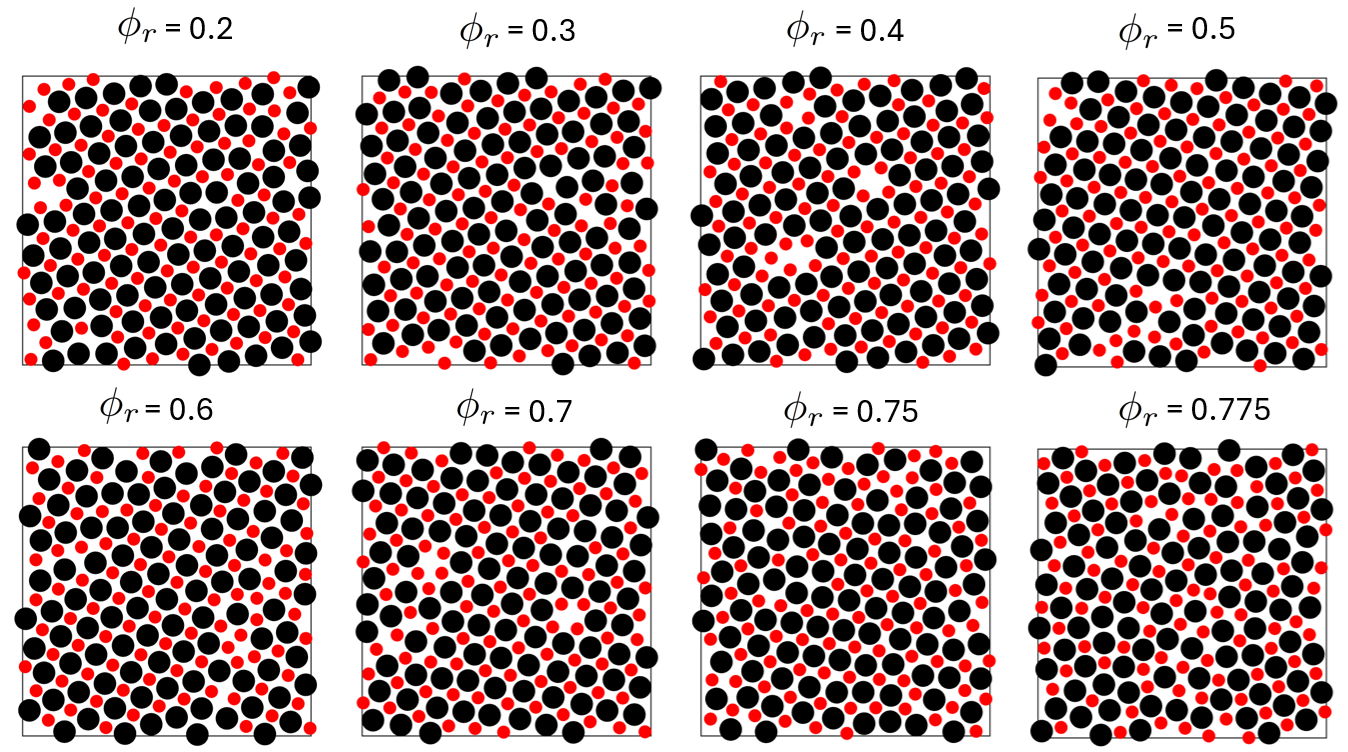}
\end{array}$
\end{center}
\caption{Final configurations of a binary system with $N=200$, a number ratio of 1:1, a size ratio 2:1, at $\phi_f \approx 0.8$ associated with varying relaxation densities from $\phi_r =$ 0.2 to 0.775} \label{Binary_QD_Effects}
\end{figure}

The relaxation density $\phi_r$, which determines the onset of the soft-shell interaction, plays a critical role in determining the system’s final structure by controlling how long-range soft interactions reshape particle arrangements. For high $\phi_r$, particles are so constrained by initial crowding associated with their hard cores that their movement slows, often leading to gridlock and preventing reorganization into the favored low-energy states. In contrast, for low $\phi_r$, soft interactions dominate, allowing particles to reorganize freely into energetically favorable (and thus, more uniform) configurations. During the re-organization, larger particles generally act as anchors that reposition first, with smaller particles filling the surrounding voids, ultimately producing an ordered, hyperuniform structure akin to a binary colloidal crystal. Figure \ref{Binary_QD_Effects} shows configurations of a binary system with number ratio 1:1 and size ratio 2:1 at $\phi_f \approx 0.8$ associated with variable relaxation densities from $\phi_r =$ 0.2 to 0.775. It can be seen that all configurations converge to expected square crystals (with defects) except for $\phi_r = 0.775$, indicating the effects of the soft-shell energy relation are very robust. In our subsequent simulations, we generally employ a small $\phi_r$ (such as 0.1) to ensure the soft-shell repulsion is fully operational to enhance the generation of hyperuniform configurations.

\subsection {Effects of the Number Ratio on Binary Systems}

\begin{figure}[ht]
\begin{center}
$\begin{array}{c}\\
\includegraphics[width=0.47\textwidth]{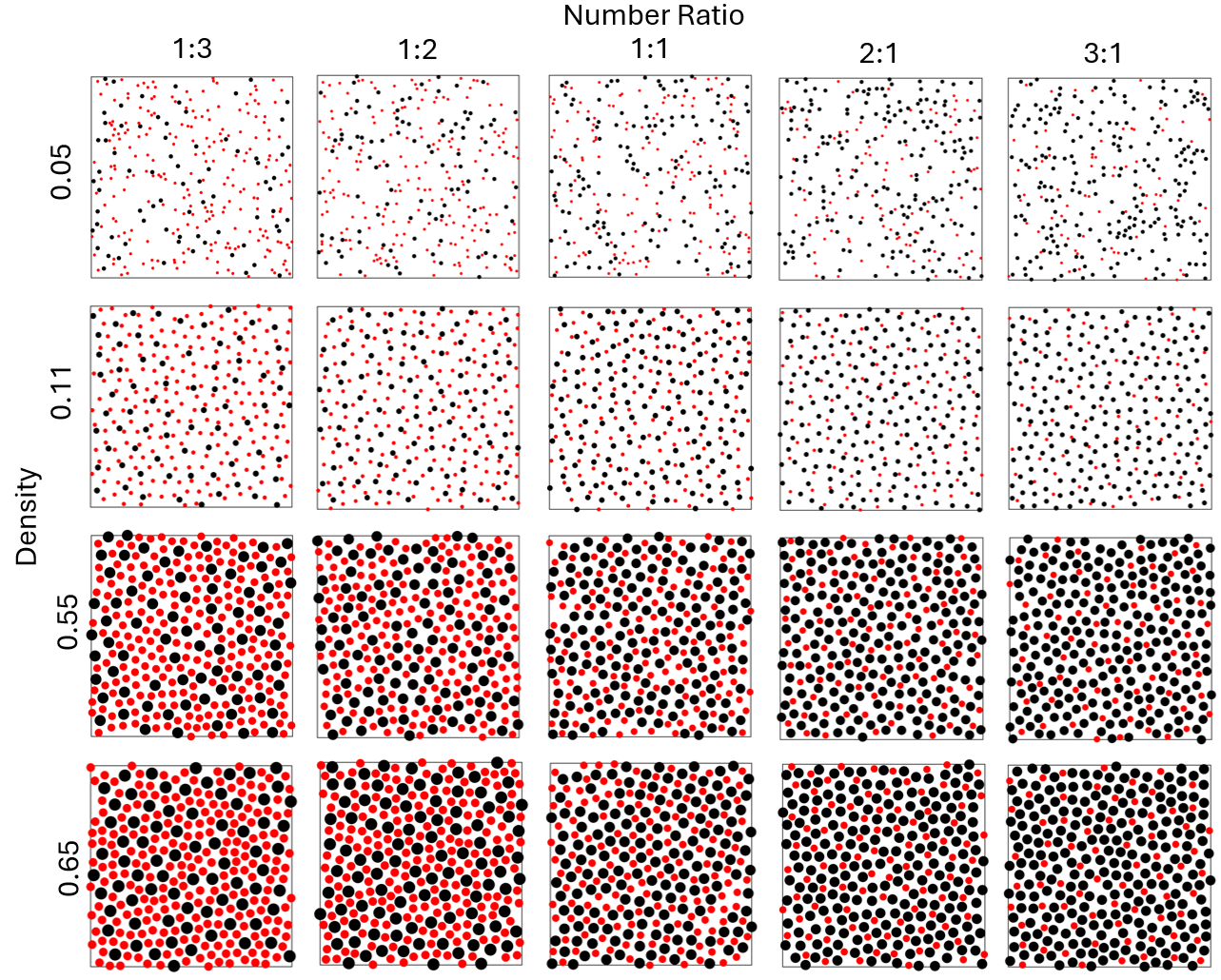}
\end{array}$
\end{center}
\caption{Configurations of binary systems with $N = 300$ particles and varying large-to-small number ratio $\gamma$ at increasing density $\phi = $0.05-0.65. The size ratio of large to small particles is 2:1.} \label{Binary_Number_Ratio}
\end{figure}

\begin{figure}[ht]
\begin{center}
$\begin{array}{c}\\
\includegraphics[width=0.475\textwidth]{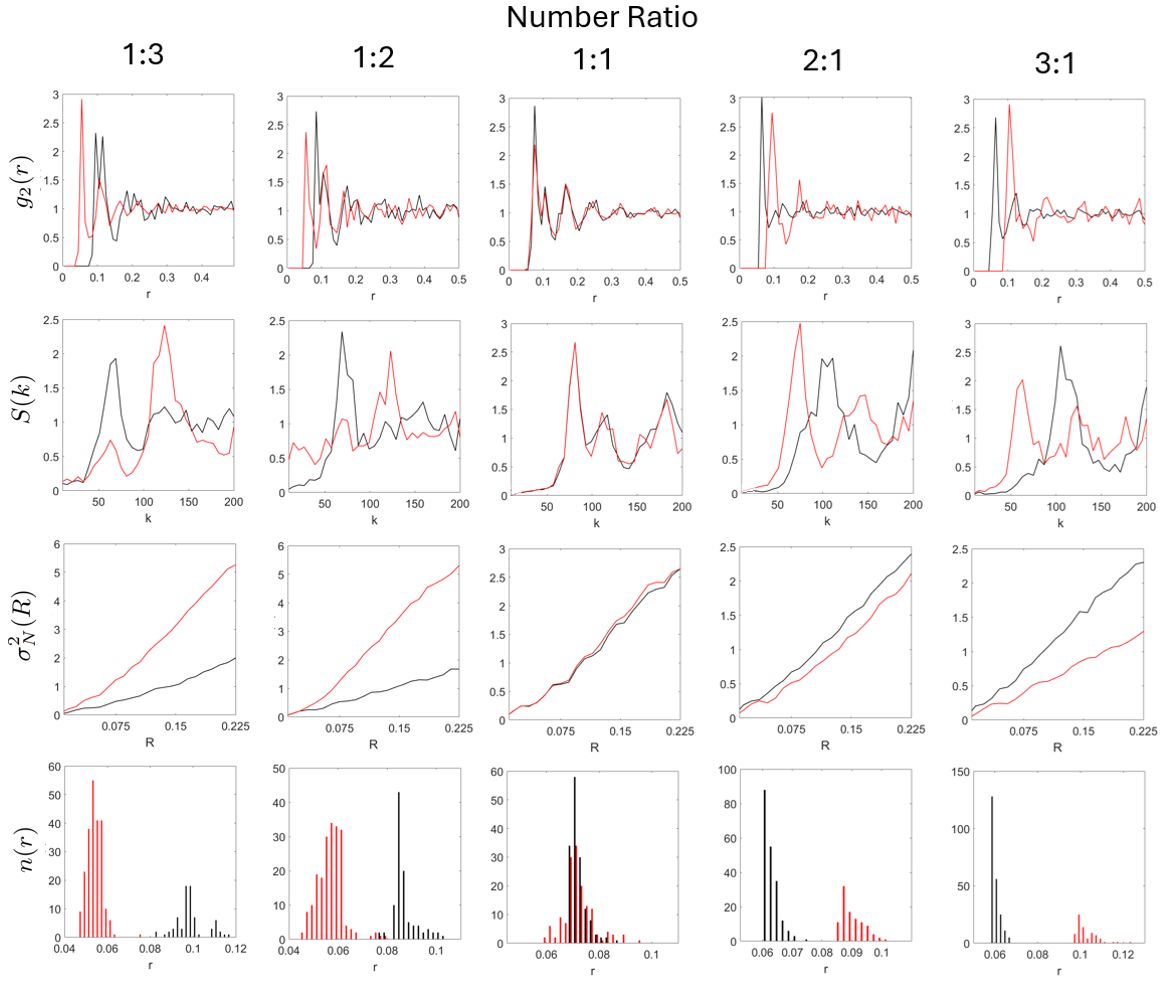}
\end{array}$
\end{center}
\caption{Statistics for $N = 300$ particle binary systems with varying large-to-small number ratios at termination density $\phi_f = 0.65$. From top to bottom, graphs correspond to the pair correlation function $g_2(r)$, static structure factor $S(k)$, number variance $\sigma^2_N(R)$, and nearest neighbor distribution $n(r)$. } \label{Binary_Number_Ratio_Stats}
\end{figure}

%Notably, for large-particle-dominated ratios (2:1, 3:1), oscillations in $g_2(r)$ are significantly dampened beyond the first peak, suggesting suppressed segregation and enhanced uniformity.

We investigate binary particle systems with varying large-to-small particle number ratios $\gamma$, spanning from small-particle-dominated (1:3) to large-particle-dominated mixtures (3:1), each relaxed at $\phi_r=0.1$ and terminated at $\phi_f=0.65$. Structural snapshots at initialization, immediately after the onset of soft-shell interactions, during intermediate growth, and at termination reveal distinct configurational pathways governed by particle number ratio. The pair correlation function $g_2(r)$ indicates short-range ordering across all systems, with majority species exhibiting pronounced first-neighbor peaks. Complementary analysis of the static structure factor $S(k)$ demonstrates that while all systems exhibit suppression at small $k$, only the balanced and large-particle-dominated systems (1:1, 2:1, and 3:1) achieve exceptionally low values of $S(k \to 0)$, with the associated small-$k$ behaviors consistent with class-I hyperuniform scaling. Number variance $\sigma^2_N(R)$ further confirms this trend: systems with balanced ratios $\gamma$ display linear growth with observation window size $R$, whereas small-particle-favored systems exhibit quadratic scaling, indicative of normal density fluctuations.

Nearest-neighbor distributions $n(r)$ provide additional insight into the configurational role of each species. In small-particle-dominated systems, the majority species exhibit strong intra-species clustering, leading to bimodal neighbor distributions characteristic of non-hyperuniform states. In contrast, large-particle-dominated systems exhibit tighter packing and enhanced correlation lengths, reflecting the greater capacity of large particles to impose structural constraints on their surroundings. Small-particle-dominated systems, however, lack the capacity to reorganize larger inclusions effectively, resulting in weaker suppression of $S(k)$ and diminished hyperuniformity. Collectively, the $g_2(r)$, $S(k)$, $\sigma^2_N(R)$, and nearest-neighbor analyses converge to the same conclusion, i.e., the balanced $\gamma = $1:1 system consistently exhibits the strongest hyperuniform signatures, with the large-particle-dominated systems (2:1 and 3:1) also performing well. These results highlight the disproportionate importance of large particles in enforcing long-range structural order and demonstrate that balanced mixtures, particularly the 1:1 configuration, provide the most favorable conditions for achieving hyperuniform states.

\begin{figure}[ht]`
\begin{center}
$\begin{array}{c}\\
\includegraphics[width=0.475\textwidth]{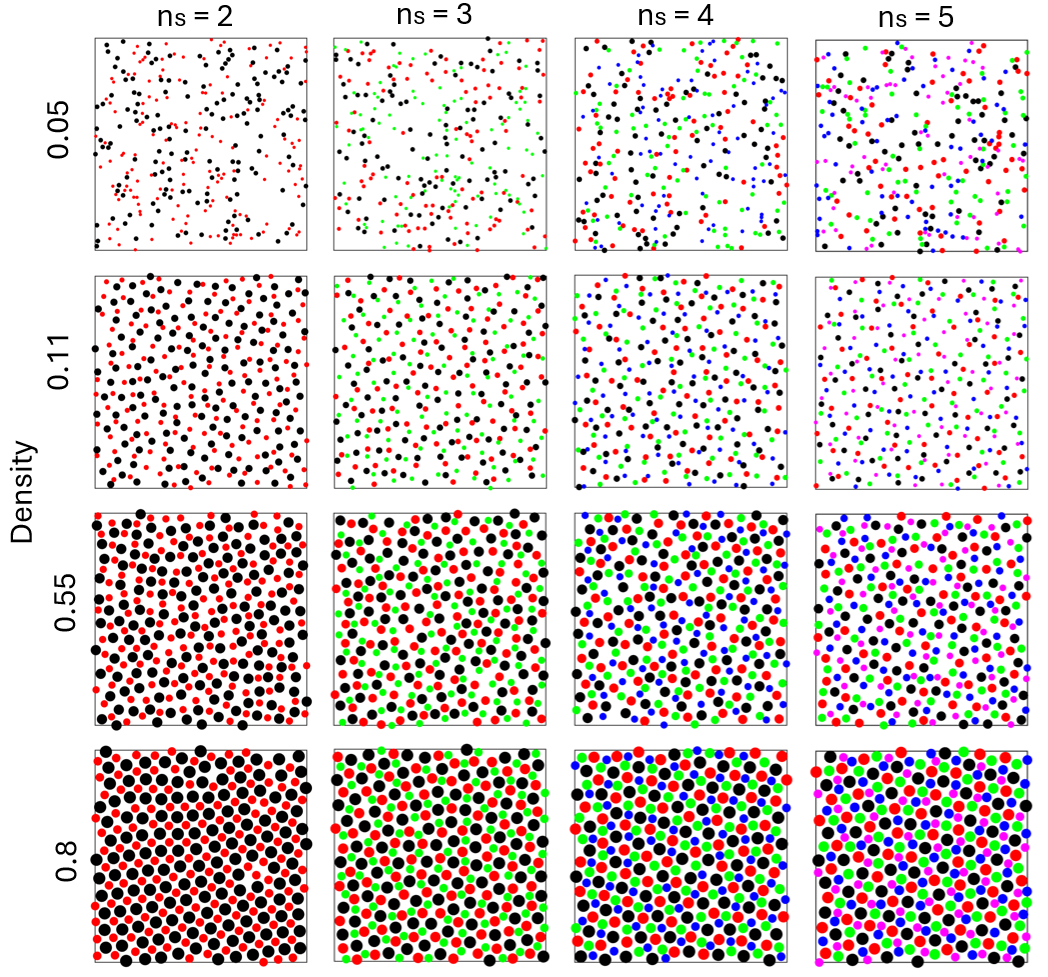}
\end{array}$
\end{center}
\caption{Configurations of multi-species systems with $N = 300$ particles and varying large-to-small number ratio $\gamma$ at increasing density $\phi = $0.05-0.8. The size ratio of largest (black) to smallest (red) particles is 2:1, with intermediate sizes assigned to  green, blue and magenta particles.} \label{Multiple_Species}
\end{figure}

\begin{figure}[ht]
\begin{center}
$\begin{array}{c}\\
\includegraphics[width=0.475\textwidth]{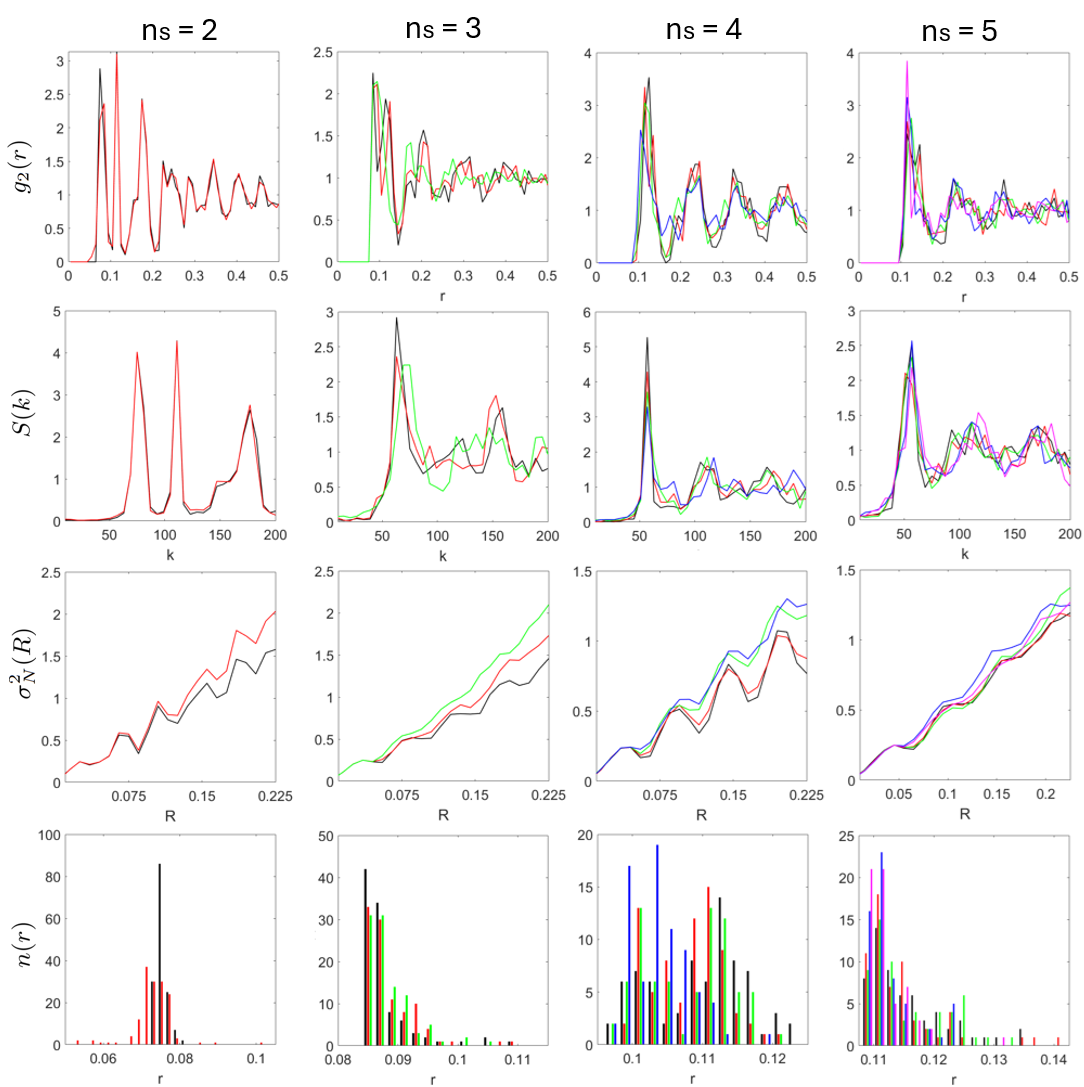}
\end{array}$
\end{center}
\caption{Statistics for multi-species systems with $N = 300$ particle and varying $n_s$ at termination density $\phi_f = 0.8$. From top to bottom, graphs correspond to the pair correlation function $g_2(r)$, static structure factor $S(k)$, number variance $\sigma^2_N(R)$, and nearest neighbor distribution $n(r)$. } \label{Multiple_Species_Stats}
\end{figure}

\subsection {Generation and Analysis of Multihyperuniform Systems}

Building on the analysis of binary systems, we next examine how increasing the number of particle species $n_s$ influences the emergence of order and hyperuniformity. Multi-species systems with $n_s = 2$, 3, 4 and 5 were initialized at a low initial density $\phi_{0}=0.05$, with particle diameters of distinct species distributed evenly between the largest particle (black, twice the size of the smallest), with intermediate sizes in descending magnitude assigned to red, green, blue and magenta particles respectively. Figure \ref{Multiple_Species} shows the snapshots of the configurations taken before and after activation of the soft-shell potential, as well as at intermediate and high densities ($\phi \approx 0.8$), which reveal that the presence of multiple species promotes efficient space filling: smaller particles consistently occupy interstitial regions, while larger particles act as anchors stabilizing the overall structure. Due to the low relaxation density $\phi_r = 0.01$ (i.e., very early onset of soft-shell interactions), the final configuration of these systems are highly defect tolerant, exhibiting suppressed segregation and instead producing dense, isotropic packings with uniform distribution of all species throughout the system.

The associated structural statistics confirm that the number of species $n_s$ can strongly modulate the resulting structural correlations. The pair correlation function $g_2(r)$ exhibits sharp oscillatory peaks for $n_s=2$, which broaden as $n_s$ increases, reflecting stronger inter-species mixing and reduced segregation. The static structure factor $S(k)$ measurements demonstrate that all systems display suppression of long-wavelength fluctuations, with particularly strong suppression in the four-species configurations. For this case ($n_s=4$), $S(k) \approx 0$ for $k< K^*$, consistent with \emph{stealthy multihyperuniformity}, where density fluctuations are eliminated across a finite range of small wavevectors, for {\it all four species} in the system. Number variance $\sigma^2_N(R)$ for large window radius $R$ for all cases indicates class-I hyperuniform scaling. In particular, the four-species configuration exhibits the lowest variance growth rate, combining strong short-range order with maximally suppressed long-range fluctuations. Thus, while increasing $n_s$ generally increases configurational disorder, the cooperative action of the soft-shell interactions and the anchoring role of larger particles enables robust disordered hyperuniform states, with the four-species case achieving a strong stealthy multihyperuniform configuration.

\subsection {Multifunctional Optical Applications}

%Pick one species for wave propogation, present the equations for the effective dielectric constant.

%Isotropic Structural Coloration, mechanical stability, light absorbing efficiency (based on average number density), and wave propogation  

%\section{Multifunctional Optical Application of Multihyperuniform Particle Composites}

\begin{figure}[ht]
\begin{center}
$\begin{array}{c}\\
\includegraphics[width=0.45\textwidth]{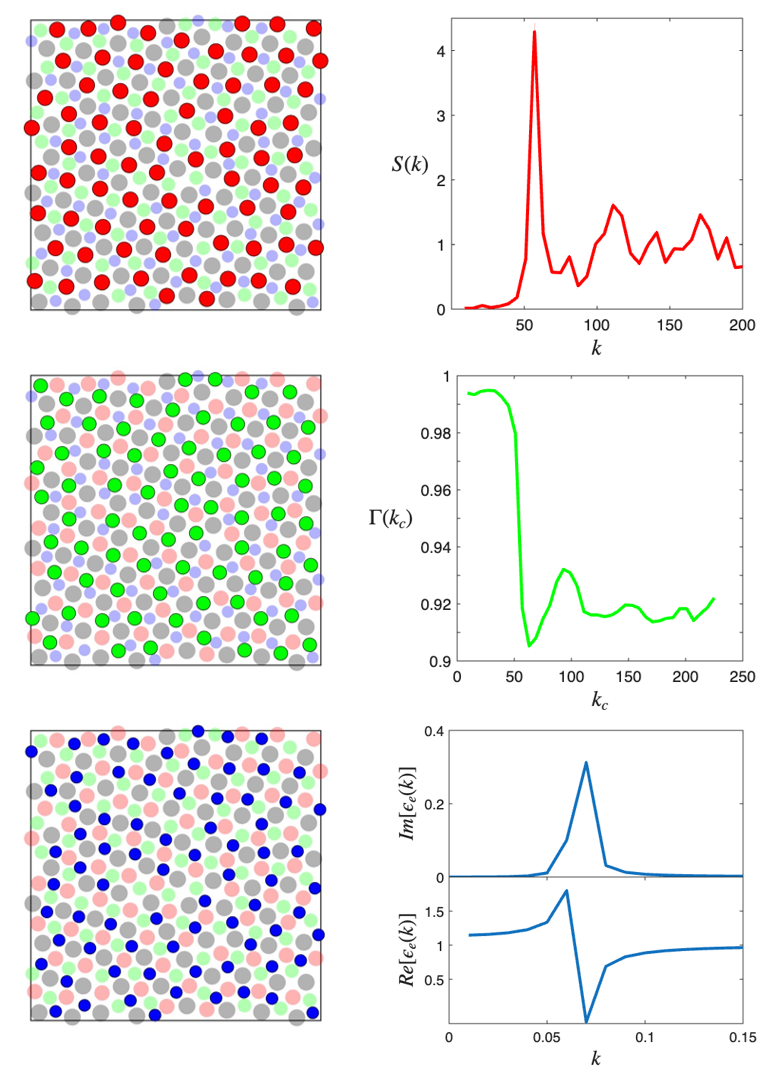}
\end{array}$
\end{center}
\caption{Illustration of a multihyperuniform particle composite, derived from the computationally generated configuration and composed of $n_s = 4$ species with distinct functionalities, simultaneously address mechanical, optical, and electromagnetic application requirements, including structural coloring, light absorbing and EM wave propagation engineering (from top to bottom).} \label{fig8}
\end{figure}

In this section, we demonstrate the potential application of the multihyperuniform particle composites, derived from the computationally generated configuration and composed of multiple particle species with distinct functionalities; these composites can offer a versatile platform for the design of advanced photonic and optoelectronic materials. For example, a four-species system can simultaneously address mechanical, optical, and electromagnetic requirements, as illustrated in Fig. \ref{fig8}. The black species is introduced primarily for reinforcement, providing mechanical stability and ensuring that the disordered composite maintains structural integrity under stress. Due to the suppressed long-wavelength density fluctuations characteristic of hyperuniformity, mechanical reinforcement is distributed isotropically, which enhances robustness without introducing preferred crystalline directions or fracture planes.

The second species (red) is engineered to produce structural coloration. In hyperuniform disordered systems, the spectral position of the strongest first peak in the structure factor $S(k)$ directly determines the dominant scattering wavelength and thus the perceived color. By tuning the size and concentration of the red particles, the position of the $S(k)$ peak can be shifted across the visible range, enabling angle-independent coloration similar to biological photonic structures such as bird feathers or beetle scales. Because the underlying arrangement is hyperuniform, the coloration remains isotropic and non-iridescent, in contrast to periodic photonic crystals that exhibit strong angle dependence.

The third species (green) is dedicated to light absorption. By introducing absorptive particles into a hyperuniform framework, photons are uniformly distributed across the medium before being absorbed, thereby reducing hotspots and enhancing the effective absorption cross section per particle. Specifically, we define the absorption efficiency factor $\Gamma(k_c)$ as 
\begin{equation}
    \Gamma(k_c) = 1 - \overline{S}(k_c)
\end{equation}
where
\begin{equation}
    \overline{S}(k_c) = \frac{d}{k_c^d}\int_0^{k_c} S(k)k^{d-1}dk
\end{equation}
and $S(k)$ is the structure factor and $k_c$ is a cut-off frequency. Fig. \ref{fig8} middle panel shows $\Gamma(k_c)$ as a function of $k_c$. It can be seen that the system leads to large $\Gamma(k_c)$, associated with less low-k density fluctuation (more hyperuniform), empirically correlated with a shorter optical mean free path $\ell_a$ and thus a higher absorption efficiency at the same particle volume fraction, thickness, and material loss. The multihyperuniform arrangement can also lead to broadband absorption with improved efficiency (more than 90 percent for all $k$ values) compared to random dispersions, since stealthy hyperuniformity reduces backscattering and prolongs photon dwell times. Such functionality is critical for applications including solar energy harvesting, photodetectors, and stealth coatings. Details of the absorptive behavior, including absorption mean free path and its correlation with diffusion spreadability, will be presented in the following section.

Finally, the blue species can be used to engineer the effective dielectric constant $\varepsilon_{e}(k_{q})$ of the composite \cite{torquato2021nonlocal}. By embedding dielectric inclusions tuned in size and fraction, one can tailor the effective permittivity of the medium to achieve desired electromagnetic propagation properties.
\begin{equation}
\frac{\varepsilon_{e}(k_{q})}{\varepsilon_{q}}
= 1 + \frac{\beta_{pq}\,\phi_{p}}
{\;\phi_{p}\left(1 - \beta_{pq}\,\phi_{p}\right)
+ \dfrac{(d-1)\pi}{2^{d/2}\Gamma(d/2)} \,
\beta_{pq}\,F\!\left(k_{q}\right)} \, ,
\end{equation}
where $\phi_p$ is the volume fraction of phase $p$, 
\begin{equation}
\beta_{pq} = \frac{\varepsilon_{p} - \varepsilon_{q}}
{\varepsilon_{p} + \varepsilon_{q}} \, .
\end{equation}
and
\begin{equation}
F(Q) = -\,\frac{2}{\pi^{2}} \int_{\mathbb{R}^{d}} 
\frac{\tilde{\chi}_{V}(\mathbf{q})}{\,|\mathbf{q}+\mathbf{Q}|^{2}-Q^{2}\,}\, d\mathbf{q} \, ,
\end{equation}
and $\tilde{\chi}_{V}(\mathbf{q})$ is the spectral density of the particle phase. As shown in Fig. \ref{fig8} bottom panel, in stealthy hyperuniform arrangements, the suppression of long-wavelength scattering ensures that electromagnetic waves propagate with minimal loss and distortion, effectively enhancing transmission through the composite for a wide range of $k$ values. This ability to simultaneously absorb, color, reinforce, and transmit establishes multihyperuniform four-species composites as multifunctional platforms for next-generation optical materials. Such designs combine isotropic structural coloration, enhanced light absorption, engineered permittivity, and mechanical stability in a single disordered yet hyperuniform system, surpassing the limitations of traditional ordered photonic crystals.

\section {conclusions and Discussion}

In this work, we have extended the study of hyperuniform materials to general multihyperuniform systems. By analyzing configurations with multiple species, varying number ratios, and interaction competitions, we demonstrated how short- and long-range correlations can be tuned to generate robust hyperuniform and stealthy hyperuniform states. These multihyperuniform arrangements naturally generalize to higher dimensions and can be realized on or off lattice, offering a versatile design space for disordered materials with tailored structural and optical properties.

The multifunctionality of multihyperuniform systems makes them especially attractive for photonics and optics. Potential applications include isotropic structural coloration, broadband absorption layers, low-loss transparent conductors, and composites with engineered dielectric constants for electromagnetic wave manipulation. By leveraging multiple species, one can simultaneously optimize distinct functionalities, for example, combining structural color with light absorption and dielectric engineering within a single composite. Such multifunctional behavior is not easily attainable in conventional periodic or random media.

Future work should focus on the systematic experimental realization of multihyperuniform composites, for instance via self-assembly of colloids, lithographic patterning, or biomimetic approaches inspired by avian photoreceptor mosaics. It is also interesting to explore the possibility of realizing multihyperuniformity in polycrystalline materials \cite{chen2016stochastic}. Theoretical advances will also be needed to establish quantitative links between structural metrics (e.g., $g_2(r)$, $S(k)$) and optical figures of merit such as absorption mean free path and effective refractive index. Ultimately, bridging design principles with scalable fabrication may enable multihyperuniform systems to serve as a new platform for robust, tunable, and multifunctional optical materials.

%\textcolor{red}{mention the advantages of physics-based method over geometry-based method}

%Disordered hyperuniformity, despite appearing random, is actually the product of large-scale order forming a cohesive spatial pattern across the entirety of the material. This pattern is formed purely by the hard-core and soft-core interactions between the different species found within the system. Same-species particles actively push away from each other when too close, leading to scenarios where they are equally distributed across the material. When combined with the hard-core interactions of other species, this leads to a system where all particles within the system are seeking equilibrium, competing to form an optimal large-scale pattern that follows their in-built restraints. Such an impetus leads to a material with the distinction of having various unique physical properties, being a mix between crystalline and disordered structure. The system suppresses large scale density fluctuations, similarly to crystal systems while also being isotropic, lacking any Braggs peaks, a property associated with amorphous or disordered materials. This dichotomy allows for previously opposing properties to be found within the same material such as the existence of large isotropic photonic band gaps.

%\section {acknowledgments}
\begin{acknowledgments}
This work was supported by the Army Research Office under Cooperative Agreement Number W911NF-22-2-0103.
\end{acknowledgments}
\smallskip

\bibliographystyle{IEEEtran}
\bibliography{reference2}

\end{document}